\documentclass{article}

\usepackage{amsmath,amssymb,amsfonts,amsthm}
\usepackage{fullpage}
\usepackage{hyperref}
\usepackage[vcentermath]{youngtab}

\newcommand{\inner}[2]{\left\langle #1 \!\mid\! #2 \right\rangle}
\newcommand{\tr}{\mathop{\mathrm{tr}}}
\newcommand{\GL}{\mathsf{GL}}
\newcommand{\R}{\mathbb{R}}
\newcommand{\C}{\mathbb{C}}
\newcommand{\Z}{\mathbb{Z}}
\newcommand{\F}{\mathbb{F}}
\newcommand{\ket}[1]{| #1 \rangle}
\newcommand{\bra}[1]{\langle #1 |}

\newcommand{\id}{\mathbf{1}}
\newcommand{\boxrep}{\beta}
\newcommand{\eps}{\epsilon}
\newcommand{\T}{\text{MM}}

\newcommand{\footremember}[2]{%
    \footnote{#2}
    \newcounter{#1}
    \setcounter{#1}{\value{footnote}}%
}
\newcommand{\footrecall}[1]{%
    \footnotemark[\value{#1}]%
} 

\title{Matrix multiplication algorithms from group orbits}

\author{Joshua A. Grochow\footremember{sfi}{Santa Fe Institute, 1399 Hyde Park Road, Santa Fe, New Mexico 87501 USA}\footnote{Department of Computer Science, University of Colorado at Boulder, 1111 Engineering Drive, ECOT 717, 430 UCB, Boulder, Colorado 80309 USA}\, and Cristopher Moore\footrecall{sfi}}

\begin{document}

\theoremstyle{definition}
\newtheorem{definition}{Definition}
\newtheorem{theorem}{Theorem}
\newtheorem{proposition}{Proposition}
\newtheorem{lemma}{Lemma}
\newtheorem{corollary}{Corollary}
\newtheorem{conjecture}{Conjecture}
\newtheorem{remark}{Remark}
\newtheorem{condition}{Condition}

\maketitle

\begin{abstract}
We show how to construct highly symmetric algorithms for matrix multiplication. In particular, we consider algorithms which decompose the matrix multiplication tensor into a sum of rank-1 tensors, where the decomposition itself consists of orbits under some finite group action.  We show how to use the representation theory of the corresponding group to derive simple constraints on the decomposition, which we solve by hand for $n=2,3,4,5$, recovering Strassen's algorithm (in a particularly symmetric form) and new algorithms for larger $n$. While these new algorithms do not improve the known upper bounds on tensor rank or the matrix multiplication exponent, they are beautiful in their own right, and we point out modifications of this idea that could plausibly lead to further improvements. Our constructions also suggest further patterns that could be mined for new algorithms, including a tantalizing connection with lattices. In particular, using lattices we give the most transparent proof to date of Strassen's algorithm; the same proof works for all $n$, to yield a decomposition with $n^3 - n + 1$ terms.
\end{abstract}

\section{Introduction}

The complexity of matrix multiplication is a central question in computational complexity, bearing on the complexity not only of most problems in linear algebra, but also of myriad combinatorial problems, e.g., various shortest path problems~\cite{zwick} and bipartite matching problems~\cite{sankowski}. The main question around matrix multiplication is whether two $n \times n$ matrices can be multiplied in time $O(n^{2 + \varepsilon})$ for every $\varepsilon > 0$; the current best upper bound on this exponent is $2.3728639$~\cite{LeGall}, narrowly beating~\cite{stothers,VW}. The best known lower bound is still only $3n^2 - o(n)$~\cite{landsbergLB}. 

Since Strassen's 1969 paper~\cite{strassen}, which showed how to beat the standard $O(n^3)$-time algorithm, it has been understood that one way to get asymptotic improvements in algorithms for matrix multiplication is to find algebraic algorithms for multiplying small, fixed-size matrices using few multiplications, and then to apply these algorithms recursively. In this note, we are thus primarily interested in algebraic algorithms for matrix multiplication. 

Recently, there have been several papers analyzing the geometry and symmetries of algebraic algorithms for small matrices~\cite{burichenkoStrassen, burichenko, landsbergRyder, landsbergMichalekLB1,CILO}. Here, we take this line of research one step further by using symmetries to discover \emph{new} algorithms for multiplying matrices of small size. While our algorithms do not improve the state-of-the-art bounds on the matrix multiplication exponent, they suggest that we can use group symmetries and group orbits to find new algorithms for matrix multiplication.  
In addition to their potential value for future endeavors, we believe that these highly symmetric matrix multiplication algorithms are beautiful in their own right, and deserve to be shared simply for their beauty.

Although this line of research was also suggested in~\cite{CILO}, who analyze the symmetries of Strassen's algorithm and promise a forthcoming paper with a new construction along these lines for $n=3$, our paper is the result of an independent project started two years ago, and we find constructions for $n=3$, $4$, and $5$.  Moreover, these constructions appear to yield a larger pattern. In the final section, we show how---although our \emph{method} of construction is more general than this---the constructions we ended up finding were in fact all instances of a single lattice-based construction yielding rank $n^3 - n + 1$ for $n \times n$ matrix multiplication. The proof that the lattice-based construction works is perhaps the simplest and most transparent proof of Strassen's algorithm that we are aware of, and the same proof works for all $n$.

\section{Background} 
\label{sec:prelim}

For general background on algebraic complexity, we refer the reader to the book~\cite{BCS}. Bl\"{a}ser's survey article~\cite{blaserSurvey}, in addition to excellent coverage around matrix multiplication, has a nice tutorial on tensors and their basic properties. 

In the algebraic setting, since matrix multiplication is a bilinear map, it is known that it can equivalently be reformulated as a tensor, and that the algebraic complexity of matrix multiplication is within a factor of 2 of the rank of this tensor. The matrix multiplication tensor for $n \times n$ matrices is 
\[
\T^{abc}_{def} = \delta^a_e \delta^b_f \delta^c_d \, , 
\]
where the indices range from $1$ to $n$.  This is also defined by the inner product
\[
\inner{\T}{A \otimes B \otimes C} = \tr ABC \, .
\]

Given vector spaces $V_1, \dotsc, V_k$, a vector $v \in V_1 \otimes \dotsb \otimes V_k$ is said to have tensor rank one if it is a separable tensor, that is, of the form $v_1 \otimes v_2 \otimes \dotsb \otimes v_k$ for some $v_i \in V_i$. The \emph{tensor rank} of $v$ is the smallest number of rank-one tensors whose sum is $v$. In the case of $\T$, we have $k = 3$ and $V_1 \cong V_2 \cong V_3 \cong \C^{n \times n}$. 

Although it won't be relevant for our results, we cannot resist mentioning border rank. The border rank of a tensor $v \in V_1 \otimes \dotsb \otimes V_k$ is the smallest $r$ such that $v$ is a limit of tensors of rank at most $r$.  The border rank is less than or equal to the rank, and indeed it can be strictly smaller.  However, the exponent of matrix multiplication is the same whether measured by tensor rank or border rank \cite{bini}.

Here is a brief summary of what is known about the tensor and border rank of small (square) matrix multiplication tensors:
\begin{table}[!htbp]
\begin{center}
\begin{tabular}{r|l|l|l|l}
$n$ & Tensor rank $\geq$ & Tensor rank $\leq$ & Border rank $\geq$ & Border rank $\leq$ \\ \hline
2 & 7~\cite{hopcroftKerr, winograd} & 7~\cite{strassen} & 7~\cite{landsberg2x2, HIL} & 7~\cite{strassen} \\
3 & 19~\cite{blaserSmall} & 23~\cite{laderman, johnsonMcloughlin, CBH} & 16~\cite{landsbergMichalekLB1} & 21~\cite{schonhage} \\
4 & 34~\cite{blaserSmall} & 49~\cite{strassen} & 29~\cite{landsbergMichalekLB1} & 49~\cite{strassen}
\end{tabular}
\end{center}
\end{table}

The tensor rank of $n \times n$ matrix multiplication is at least $3n^2 - o(n)$~\cite{landsbergLB}. The previous best bound was $(5/2) n^2 - 3n$~\cite{blaserLB}, and the $3n^2 - o(n)$ beats this out for $n > 84$. The border rank of $n \times n$ matrix multiplication is at least $2n^2 - \lceil \log_2 n \rceil - 1 $~\cite{landsbergMichalekLB2} (and at least $2n^2 - n + 1$~\cite{landsbergMichalekLB1}, which is a tighter lower bound for all $n < 6$). 

The current world record algorithm~\cite{LeGall} and its immediate precursors~\cite{stothers, VW} were achieved by analyzing powers of the Coppersmith--Winograd tensor~\cite{CW}. Ambainis, Filmus, and Le Gall~\cite{AFL} showed that taking further powers of the Coppersmith--Winograd tensor cannot get an exponent better than $2.3725$, and that even certain generalized versions of this method cannot beat exponent $2.3078$. The smallest square cases which could beat these bounds (world record, $2.3725$, and $2.3078$, respectively), based on the lower bounds above, are:
\begin{itemize}
\item Showing the border rank of $6 \times 6$ matrix multiplication is at most 70 would beat the world record (and even exponent $2.3725$). This must be border rank, as the above bounds show that its rank is at least 72. (Its border rank is at least 68.)

\item Showing the tensor rank of $8 \times 8$ matrix multiplication is at most 138 would beat the world record (and $2.3725$). Its rank is at least 136, and the current best upper bound is $7^3 = 343$ (the tensor cube of Strassen's algorithm).

\item Showing the border rank of $9 \times 9$ matrix multiplication is at most 159 would beat the stronger bound of $2.3078$. This must be border rank, as its rank is at least 176. (Its border rank is at least 157.)

\item Showing that the tensor rank of $16 \times 16$ matrix multiplication is at most 600 would beat the stronger bound of $2.3078$. Its rank is at least 592, and the current best upper bound is $7^4 = 2401$ (the fourth tensor power of Strassen's algorithm).
\end{itemize}


\section{Using Fourier analysis to rediscover Strassen as a group orbit} 
\label{sec:strassen}

We begin by ``Fourier analyzing'' the matrix multiplication tensor.  First recall that $\T$ is defined by its symmetries (see, e.g.,~\cite{burgisserIkenmeyer}), that is, up to a constant, it is the unique operator fixed under the following action of $\GL(n)^3$. For, given $X, Y, Z \in \GL(n)$, we have
\begin{equation}
\label{eq:action}
\T = (X \otimes Y \otimes Z) \T (Z^{-1} \otimes X^{-1} \otimes Y^{-1}) \, , 
\end{equation}
where hopefully the notation is clear.  To see that $\T$ has this symmetry, note that 
\[
\tr ABC = \tr (Z^{-1} A X)(X^{-1} B Y)(Y^{-1} C Z) \, .
\]
To see that $\T$ is the only such operator up to a constant, we simply note that if $\rho \cong \C^n$ is the standard representation of $\GL(n)$, then $\rho^{\otimes 3}$ is an irreducible representation of $\GL(n)^3$.  Taking the tensor product with its dual $(\rho^{\otimes 3})^*$ then gives a single copy of the trivial representation, so $\T$ lives in---and therefore spans---this one-dimensional subspace. 

The same argument applies to any group $G$ that has an $n$-dimensional irreducible presentation, which we will also call $\rho$.  Since $\rho^{\otimes 3}$ is an irrep of $G^3$, $\rho^{\otimes 3} \otimes (\rho^{\otimes 3})^*$ contains a unique one-dimensional subspace of operators invariant under the action of $G^3$ where $X, Y, Z$ are each $\rho(g)$ for some $g \in G$. This suggests that a good way to search for new matrix multiplication algorithms is to start with sums of separable tensors where the sum has some symmetry built in from the beginning.

We can also consider the action of the diagonal subgroup $G = \{(g,g,g) : g \in G\} \subset G^3$, which conjugates each factor in the tensor product by the same group element.  Clearly $\T$ is invariant under this action, but it is no longer the only such operator, since $\rho^{\otimes 3}$ now has multiple copies of the trivial representation when acted on by this smaller group.  The idea, independently suggested in \cite{CILO}, is to study how $\T$ lies across these various trivial subspaces.  In this section we carry out this program in the case $n=2$ and $G=S_3$, where $\rho$ is the standard irrep of $S_3$.  

The conjugation action of $S_3$ on the space of $2 \times 2$ matrices is $\rho \otimes \rho^* \cong \id \oplus \pi \oplus \rho$ where $\id$ and $\pi$ are the trivial and parity representations respectively.  To make things more concrete, we choose a basis for $\rho$, which corresponds to the rotations and reflections of an equilateral triangle centered on the origin, whose corners labeled $1, 2, 3$ lie at $(1,0)$, $(-1/2,\sqrt{3}/2)$, and $(-1/2,-\sqrt{3}/2)$.  Then the generators of the group correspond to the unitary matrices
\begin{equation}
\label{eq:sigma}
\rho((23)) = \begin{pmatrix} 1 & 0 \\ 0 & -1 \end{pmatrix} \triangleq \phi
\quad \text{and} \quad
\rho((123)) = \begin{pmatrix} -1/2 & -\sqrt{3}/2 \\ \sqrt{3}/2 & -1/2 \end{pmatrix} \triangleq \sigma \, ,
\end{equation}
where we think of these as acting on column vectors on the left.  Geometrically, $\phi$ flips the plane around the $x$-axis, and $\sigma$ rotates it $2\pi/3$ counterclockwise.  

The subspaces corresponding to $\id$, $\pi$, and $\rho$ are spanned by the following vectors in $\rho \otimes \rho^*$, which we denote by $\dim \rho \times \dim \rho^* = 2 \times 2$ matrices:
\[
\id = \begin{pmatrix} 1 & 0 \\ 0 & 1 \end{pmatrix} \quad
\pi = \begin{pmatrix} 0 & -1 \\ 1 & 0 \end{pmatrix} \quad
\rho_x = \begin{pmatrix} 1 & 0 \\ 0 & -1 \end{pmatrix} \quad 
\rho_y = \begin{pmatrix} 0 & -1 \\ -1 & 0 \end{pmatrix} \, , 
\]
where we abuse notation by identifying the irreps $\id$ and $\pi$ with matrices that span them.  The reader can check that 
\begin{equation}
\label{eq:identities-pi}
\phi \pi \phi^\dagger = -\pi 
\quad \text{and} \quad 
\sigma \pi \sigma^{-1} = \pi \, ,  
\end{equation}
confiming that $\pi$ spans the parity representation.  We have chosen $\rho_x$ and $\rho_y$ so that they correspond to the basis vectors $(1,0)$ and $(0,1)$ in the original basis for $\rho$.  In particular, 
\begin{equation}
\label{eq:identities-rho}
\phi \rho_x \phi^{\dagger} = \rho_x \, , \quad
\phi \rho_y \phi^{\dagger} = -\rho_y \, , \quad
\sigma \rho_x \sigma^{\dagger} = -\frac{1}{2} \rho_x + \frac{\sqrt{3}}{2} \rho_y \, , \quad
\text{and} \quad
\sigma \rho_y \sigma^{\dagger} = -\frac{\sqrt{3}}{2} \rho_x - \frac{1}{2} \rho_y \, ,
\end{equation}
(We know that $\rho_x = \phi$, but we want to distinguish $\rho_x$, which is a vector in $\rho \otimes \rho^*$, from $\phi$, which is an operator acting on that space by conjugation.)  

We now have a convenient basis $\{\id, \pi, \rho_x, \rho_y\}^{\otimes 3}$ for the space of triples of $2 \times 2$ matrices, i.e., for $(\rho \otimes \rho^*)^{\otimes 3}$.  Next we write $\T$ in this basis.  Since these $64$ basis vectors are orthogonal (under the unique $S_3$-invariant inner product of $A, B \in \rho \otimes \rho^*$ defined as $\inner{A}{B} = \tr(A^\dagger B)$), we have 
\[
\T = \sum_{(\alpha, \boxrep, \gamma) \in \{\id, \pi, \rho_x, \rho_y\}^3}
c_{\alpha,\boxrep,\gamma} \,\alpha \otimes \boxrep \otimes \gamma \, , 
\]
where
\[
c_{\alpha,\boxrep,\gamma} 
= \frac{\inner{\T}{\alpha \otimes \boxrep \otimes \gamma}}{\inner{\alpha \otimes \boxrep \otimes \gamma}{\alpha \otimes \boxrep \otimes \gamma}}
= \frac{1}{8} \tr \alpha \boxrep \gamma \, . 
\]
(The $1/8$ here comes from the fact that $\inner{\alpha \otimes \boxrep \otimes \gamma}{\alpha \otimes \boxrep \otimes \gamma} = \inner{\alpha}{\alpha} \inner{\boxrep}{\boxrep} \inner{\gamma}{\gamma}$, and for all $\alpha \in \{\id, \pi, \rho_x, \rho_y\}$ we have $\inner{\alpha}{\alpha} = 2$.)
This gives 
\begin{align}
\T = \frac{1}{4} \bigl( & \id \otimes \id \otimes \id  \\
-\;& \id \otimes \pi \otimes \pi  
- \pi \otimes \id \otimes \pi 
- \pi \otimes \pi \otimes \id 
\nonumber \\
+\;& \id \otimes \rho_x \otimes \rho_x 
+ \rho_x \otimes \id \otimes \rho_x 
+ \rho_x \otimes \rho_x \otimes \id 
\nonumber \\
+\;& \id \otimes \rho_y \otimes \rho_y 
+ \rho_y \otimes \id \otimes \rho_y
+ \rho_y \otimes \rho_y \otimes \id 
\nonumber \\
+\;& \pi \otimes \rho_x \otimes \rho_y
+ \rho_y \otimes \pi \otimes \rho_x 
+ \rho_x \otimes \rho_y \otimes \pi 
\nonumber \\
-\;& \pi \otimes \rho_y \otimes \rho_x
- \rho_x \otimes \pi \otimes \rho_y 
- \rho_y \otimes \rho_x \otimes \pi \bigr) \, .
\label{eq:t-fourier}
\end{align}

We streamlined this calculation by noting that each of the $27$ tensor products of representations $\{\id, \pi, \rho\}^{\otimes 3}$ has at most one copy of the trivial representation of the diagonal $S_3$ action, and that it is only these triples for which we can have $c_{\alpha,\boxrep,\gamma} \ne 0$.  In particular, the third and fourth lines correspond to the trivial subspace of $\id \otimes \rho \otimes \rho$, which is spanned by 
\[
\id \otimes \psi^{\rho \rho}_\id 
\quad \text{where} \quad
\psi^{\rho \rho}_\id = \rho_x \otimes \rho_x + \rho_y \otimes \rho_y \, ,
\]
since $\psi^{\rho \rho}_\id$ spans the trivial subspace of $\rho \otimes \rho$.  Similarly, the fifth and sixth lines correspond to the trivial subspace of $\pi \otimes \rho \otimes \rho$, which is spanned by 
\[
\pi \otimes \psi^{\rho\rho}_\pi 
\quad \text{where} \quad
\psi^{\rho \rho}_\pi = \rho_x \otimes \rho_y - \rho_y \otimes \rho_x \, , 
\]
since $\psi^{\rho \rho}_\pi$ spans the parity subspace of $\rho \otimes \rho$.  
Interestingly, $\rho \otimes \rho \otimes \rho$ does contain a copy of the trivial representation, spanned by 
\begin{equation}
\label{eq:rhorhorho}
\psi^{\rho\rho\rho}_\id 
= \rho_x \otimes \rho_x \otimes \rho_x 
- \rho_x \otimes \rho_y \otimes \rho_y 
- \rho_y \otimes \rho_x \otimes \rho_y 
- \rho_y \otimes \rho_y \otimes \rho_x 
= \frac{(\rho_x + i \rho_y)^{\otimes 3} + (\rho_x - i \rho_y)^{\otimes 3}}{2} \, . 
\end{equation}
However, the corresponding coefficients in $\T$ are all zero, since the product of any three $\rho \in \{\rho_x,\rho_y\}$ has zero trace.

While some of its terms can be combined, taking~\eqref{eq:t-fourier} directly does not lead to a low tensor rank for $\T$.  We now set about retroactively discovering Strassen's algorithm, with the following ansatz:
\begin{align}
\T 
&= \id \otimes \id \otimes \id 
+ \sum_{g \in S_3} (\rho(g) m_1 \rho(g^{-1})) \otimes (\rho(g) m_2 \rho(g^{-1})) \otimes (\rho(g) m_3 \rho(g^{-1})) 
\nonumber \\
&= \id \otimes \id \otimes \id 
+ \sum_{g \in S_3} \rho(g)^{\otimes 3} (m_1 \otimes m_2 \otimes m_3) \rho(g)^{\dagger \otimes 3}
\, .
\label{eq:ansatz}
\end{align}
That is, besides the term $\id^{\otimes 3}$ which we generously grant ourselves, we assume that the other six terms consist of the orbit of some rank-1 tensor $m_1 \otimes m_2 \otimes m_3$ under the diagonal conjugation action of $S_3$.  Moreover, we assume that $m_1, m_2, m_3$ form an orbit under the order-3 element $(1,2,3)$.  Since $\sigma = \rho((123))$, this gives
\begin{equation}
\label{eq:z3orbit}
m_1 = m \, , \quad m_2 = \sigma m \sigma^\dagger \, , \quad m_3 = \sigma^2 m (\sigma^\dagger)^2 \, .
\end{equation}
Note that this ansatz is manifestly invariant under both the diagonal conjugation action and the $\Z_3$ symmetry which rotates the three factors of the tensor product.

Now, since summing over an orbit annihilates any non-trivial irreps of $S_3$, it suffices for $m_1 \otimes m_2 \otimes m_3$ to have the correct projection into each copy of the trivial representation, which we enumerated above.  
This gives the following necessary and sufficient conditions:
\begin{align*}
6 \,\frac{\inner{m_1 \otimes m_2 \otimes m_3}{\id \otimes \id \otimes \id}} 
{\inner{\id \otimes \id \otimes \id}{\id \otimes \id \otimes \id}} &= \frac{1}{4} - 1 \\
6 \,\frac{\inner{m_1 \otimes m_2 \otimes m_3}{\id \otimes \pi \otimes \pi}} 
{\inner{\id \otimes \pi \otimes \pi}{\id \otimes \pi \otimes \pi}} &= -\frac{1}{4} \\
6 \,\frac{\inner{m_1 \otimes m_2 \otimes m_3}{\id \otimes \psi^{\rho\rho}_\id}}
{\inner{\id \otimes \psi^{\rho\rho}_\id}{\id \otimes \psi^{\rho\rho}_\id}} &= \frac{1}{4} \\
6 \,\frac{\inner{m_1 \otimes m_2 \otimes m_3}{\pi \otimes \psi^{\rho\rho}_\pi}}
{\inner{\pi \otimes \psi^{\rho\rho}_\pi}{\pi \otimes \psi^{\rho\rho}_\pi}} &= \frac{1}{4} \\
6 \,\frac{\inner{m_1 \otimes m_2 \otimes m_3}{\psi^{\rho\rho\rho}_\id}}
{\inner{\psi^{\rho\rho\rho}_\id}{\psi^{\rho\rho\rho}_\id}} &= 0
 \, , 
\end{align*}
or more concretely (where we use $\pi^\dagger = -\pi$, $\rho_x^\dagger = \rho_x$, and $\rho_y^\dagger = \rho_y$)
\begin{gather*}
(\tr m_1)(\tr m_2)(\tr m_3) = -1 \\
(\tr m_1)(\tr \pi m_2)(\tr \pi m_3) = -\frac{1}{3} \\
(\tr m_1)\big( (\tr \rho_x m_2)(\tr \rho_x m_3) + (\tr \rho_y m_2)(\tr \rho_y m_3) \big) = \frac{2}{3} \\
(\tr \pi m_1)\big( (\tr \rho_x m_2)(\tr \rho_y m_3) - (\tr \rho_y m_2)(\tr \rho_x m_3) \big) = -\frac{2}{3} \\
\big( \tr (\rho_x + i \rho_y) m_1 \bigr) \bigl(\tr (\rho_x + i \rho_y) m_2 \bigr) \bigl(\tr (\rho_x + i \rho_y) m_3 \bigr) 
+ \big( \tr (\rho_x - i \rho_y) m_1 \bigr) \bigl(\tr (\rho_x - i \rho_y) m_2 \bigr) \bigl(\tr (\rho_x - i \rho_y) m_3 \bigr)
= 0 \, . 
\end{gather*}
Note that the rotated versions of these equations hold automatically given our assumption~\eqref{eq:z3orbit}, since these are trivial subspaces under the diagonal conjugation action and, e.\,g., $m_2 \otimes m_3 \otimes m_1 = \sigma^{\otimes 3} (m_1 \otimes m_2 \otimes m_3) \sigma^{\dagger \otimes 3}$.  Moreover, using~\eqref{eq:identities-pi} and~\eqref{eq:identities-rho} and a few lines of algebra, the above equations simplify to
\begin{gather}
(\tr m)^3 = -1 \label{eq:1} \\
(\tr m)(\tr \pi m)^2 = -\frac{1}{3} \label{eq:2} \\
(\tr m) \times -\frac{1}{2} \big( (\tr \rho_x m)^2 + (\tr \rho_y m)^2 \big) = \frac{2}{3} \label{eq:3} \\
(\tr \pi m) \times -\frac{\sqrt{3}}{2} \big( (\tr \rho_x m)^2 + (\tr \rho_y m)^2 \big) = -\frac{2}{3} \label{eq:4} \\
(\tr \rho_x m)^3 - 3 (\tr \rho_x m)(\tr \rho_y m)^2 = 0 \label{eq:finally} \, . 
\end{gather}
Since we can multiply $m$ by any cube root of $1$ without changing $m_1 \otimes m_2 \otimes m_3$, we are free to solve~\eqref{eq:1} by setting 
\begin{equation}
\label{eq:tru}
\tr m = -1 \, .
\end{equation}
Then~\eqref{eq:3} and~\eqref{eq:4} imply 
\begin{equation}
\label{eq:trpiu}
\tr \pi m = \frac{1}{\sqrt{3}} \, , 
\end{equation}
which solves~\eqref{eq:2} and gives
\begin{equation}
(\tr \rho_x m)^2 + (\tr \rho_y m)^2 = \frac{4}{3} \label{eq:5} \, . 
\end{equation}

If we write
\[
m = \begin{pmatrix} m_{11} & m_{12} \\ m_{21} & m_{22} \end{pmatrix} \, , 
\]
then
\begin{equation}
\label{eq:det1}
(\tr m)^2 + (\tr \pi m)^2 
= (m_{11}+m_{22})^2 + (m_{21}-m_{12})^2 
= |m|^2 + 2 \det m
\end{equation}
where $|m|^2 = \tr m^\dagger m$ denotes the Frobenius norm, while
\begin{equation}
\label{eq:det2}
(\tr \rho_x m)^2 + (\tr \rho_y m)^2 
= (m_{11} - m_{22})^2 + (m_{12} + m_{21})^2 
= |m|^2 - 2 \det m \, . 
\end{equation}
But~\eqref{eq:tru}, \eqref{eq:trpiu}, and~\eqref{eq:5} give
\[
(\tr m)^2 + (\tr \pi m)^2 = (\tr \rho_x m)^2 + (\tr \rho_y m)^2 = \frac{4}{3} \, ,
\]
so we have 
\[
\det m = 0 
\quad \text{and} \quad
|m|^2 = \frac{4}{3} \, . 
\]

It follows that $m$ has rank $1$, and that it can be written $m = \ket{u} \bra{v}$ for some vectors $u, v$ such that 
\[
\tr m = \inner{v}{u} = -1 \, , 
\quad
|m|^2 = |u|^2 |v|^2 = \frac{4}{3} \, , 
\quad \text{and} \quad
\tr \pi m = \inner{v}{\pi u} = \frac{1}{\sqrt{3}} \, .  
\]
Since rescaling $u$ and $v$ to $cu$ and $c^{-1}v$ preserves $\ket{u} \bra{v}$ for any constant $c$, we can assume that $|u|=1$ and $|v| = 2/\sqrt{3}$.  In addition, the first two of these conditions imply that the angle between $u$ and $v$ has cosine $\inner{v}{u}/(|u||v|) = -\sqrt{3}/2$, and since $\pi$ rotates the plane counterclockwise by $\pi/2$ the third condition implies that this angle is $5\pi/6$.  Thus we can write
\begin{equation}
\label{eq:u-v}
u = (\cos \theta, \sin \theta) 
\quad \text{and} \quad
v = \frac{2}{\sqrt{3}} \left(\cos (\theta + 5\pi/6), \sin (\theta + 5\pi/6) \right) 
= \frac{2}{3} (\sigma u - u) \, . 
\end{equation}
We can show that $\theta$ belongs to a finite set using~\eqref{eq:finally}, which is the only one of our Fourier-based equations we have not used so far.  Substituting $m = \ket{u} \bra{v}$ yields
\begin{equation}
\label{eq:finally-uv}
\inner{v}{\rho_x u}^3 - 3 \inner{v}{\rho_x u} \inner{v}{\rho_y u}^2 
= -\frac{8}{3\sqrt{3}} \sin 6\theta 
= 0 \, , 
\end{equation}
so $\theta$ is a integer multiple of $\pi/6$, and any such multiple yields a decomposition of $\T$.  In particular, setting $\theta = 0$ gives
\begin{equation}
\label{eq:uv-strassen}
u=(1,0)
\quad \text{and} \quad
v=(-1,\sqrt{3}) \, , 
\end{equation}
in which case
\[
m = \begin{pmatrix}
-1 & \sqrt{3} \\ 0 & 0 
\end{pmatrix} \, .
\]

Finally, we note that if we consider the following set of unit vectors which form an orbit under $S_3$, 
\[
w_1 = (1,0) \, , \quad
w_2 = \left( -\frac{1}{2}, \frac{\sqrt{3}}{2} \right) \, , \quad 
w_3 = \left( -\frac{1}{2}, -\frac{\sqrt{3}}{2} \right) \, , \quad 
\]
then we can express $u$ and $v$ in~\eqref{eq:uv-strassen} as 
\[
u = w_1 
\quad \text{and} \quad
v = \frac{2}{3} (w_2 - w_1) \, . 
\]
As we will see below, a similar construction works for the next few values of $n$.

\section{Constructions for larger $n$}

Strassen's algorithm, or rather its symmetric form derived above, suggests a family of matrix multiplication algorithms.  Namely, suppose we have a group $G$ with an irreducible representation $\rho$ of dimension $n$ (which we assume for simplicity is real) and which contains an order-3 element $\sigma$.  Then we consider decompositions of the matrix multiplication tensor consisting of $\id^{\otimes 3}$ plus an orbit of some rank-1 matrix $m=\ket{u} \bra{v}$,
\begin{align}
\T 
&= \id \otimes \id \otimes \id 
+ \sum_{g \in G} \rho(g)^{\otimes 3} 
\left( m \otimes \sigma m \sigma^{-1} \otimes \sigma^2 m \sigma^{\dagger 2} \right) 
\rho(g)^{\dagger \otimes 3} \nonumber \\ 
&= \id \otimes \id \otimes \id 
+ \sum_{g \in G} \rho(g)^{\otimes 3} 
\left( \ket{u} \bra{v} \otimes \ket{\sigma u} \bra{\sigma v} \otimes \ket{\sigma^2 u} \bra{\sigma^2 v} \right) 
\rho(g)^{\dagger \otimes 3}
\label{eq:family}
\end{align}
where $u, v$ are vectors in $\R^n$ and $\sigma = \rho(h)$.  Note that any decomposition of this form is manifestly symmetric under the diagonal conjugation action of $G$, as well as the $\Z_3$ action that rotates the three factors in the tensor product; both these actions simply permute the terms in the sum.  The question is whether there are $u, v$ such that~\eqref{eq:family} is also symmetric under the $G^3$ action, in which case it is a valid decomposition of the matrix multiplication tensor.

The constructions we have found have $|G| = n^3-n$, giving a decomposition of tensor rank $n^3-n+1$.  For $n > 2$ this is not optimal: it yields decompositions of rank $25$ and $61$ for $n=3$ and $n=4$ respectively, which exceed the known upper bounds of $23$~\cite{laderman, johnsonMcloughlin, CBH} and $49$~\cite{strassen}.  However, we hope that similar constructions involving multiple orbits will yield optimal decompositions in the future.  For instance, the tensor square of Strassen's algorithm can be described in terms of the four-dimensional irrep of $S_3^2$, with one orbit of $36$ rank-1 matrices, two orbits of $6$ rank-2 matrices, and $\id^{\otimes 3}$, for a total tensor rank of $49$.

We start by writing some necessary conditions on $u$ and $v$.  Viewed as a linear operator on $(\R^n)^{\otimes 3}$, the matrix multiplication tensor has trace $\tr \T = n$ and Frobenius norm $|\T|^2 = \inner{\T}{\T} = n^3$.  Finally, we have $\tr \T^2 = \tr \T = \inner{\T}{\T^\dagger} = n$.  Since the trace is basis-invariant, the traces of the terms in~\eqref{eq:family} are
\begin{gather*}
\tr \id^{\otimes 3} = n^3 \\
\tr \left( \ket{u} \bra{v} \otimes \ket{\sigma u} \bra{\sigma v} \otimes \ket{\sigma^2 u} \bra{\sigma^2 v} \right)
= \left( \tr \ket{u} \bra{v} \right)^3 
= \inner{v}{u}^3 \, , 
\end{gather*}
and the inner product of each term with $\T$ is 
\begin{gather*}
\inner{\T}{\id^{\otimes 3}} = \tr \id = n \\
\inner{\T}{\ket{u} \bra{v} \otimes \ket{\sigma u} \bra{\sigma v} \otimes \ket{\sigma^2 u} \bra{\sigma^2 v}}
= \tr \left( \ket{u} \inner{v}{\sigma u} \inner{\sigma v}{\sigma^2 u} \bra{\sigma^2 v} \right) 
= \inner{v}{\sigma u}^3 \, ,
\end{gather*}
where we have used the fact that $\sigma$ is unitary.
Finally we have
\begin{gather*}
\inner{\T}{\id^{\dagger \otimes 3}} = \tr \id = n \\
\inner{\T}{\left( \ket{u} \bra{v} \otimes \ket{\sigma u} \bra{\sigma v} \otimes \ket{\sigma^2 u} \bra{\sigma^2 v} \right)^\dagger} 
= \tr \left( \ket{v} \inner{u}{\sigma v} \inner{\sigma u}{\sigma^2 v} \bra{\sigma^2 u} \right) 
= \inner{v}{\sigma^2 u}^3 \, .
\end{gather*}
Thus~\eqref{eq:family} gives
\begin{equation}
\label{eq:un-necessary0}
n = n^3 + |G| \inner{v}{u}^3 \, , \quad
n^3 = n + |G| \inner{v}{\sigma u}^3 \, , \quad 
\text{and} \quad
n = n + \inner{v}{\sigma^2 u}^3 \, . 
\end{equation}
As before, we use our freedom to multiply, say, $u$ by a cube root of unity.  Along with $|G| = n^3-n$, this gives the conditions
\begin{equation}
\label{eq:uv-necessary}
\inner{v}{u} = -1 \, , \quad
\inner{v}{\sigma u} = +1 \, , \quad
\inner{v}{\sigma^2 u} = 0 \, .
\end{equation}
For instance, suppose $n=2$, $\rho$ is the standard representation of $G=S_3$, and $u$ is a basis vector $(1,0)$.  If $\sigma$ rotates $\R^2$ by $2 \pi / 3$ counterclockwise, the unique vector $v$ that satisfies~\eqref{eq:uv-necessary} is $(-1,1/\sqrt{3})$, recovering the pair $u,v$ we constructed for Strassen's algorithm~\eqref{eq:uv-strassen} in the previous section.  

More generally, suppose an order-3 element $\sigma$ rotates a two-dimensional subspace $R_\sigma$ by $2 \pi / 3$, and fixes its $(n-2)$-dimensional orthogonal complement $\overline{R_\sigma}$.  We know from~\eqref{eq:uv-necessary} that $u \ne \sigma u$, so $u$ has a nonzero projection into $R_\sigma$.  Since we only care about the product $m=\ket{u} \bra{v}$, we can rescale $u$ so that this projection is of unit length; denote it by $y$.  We can then write $u = y + u_\perp$ where $u_\perp \in \overline{R_\sigma}$.  Similarly write $v = z + v_\perp$ where $z \in R_\sigma$ and $v_\perp \in \overline{R_\sigma}$.  

Since $u+\sigma u+\sigma^2 u = 3 u_\perp$, summing the three conditions of~\eqref{eq:uv-necessary} implies that $\inner{v}{u_\perp} = \inner{v_\perp}{u_\perp} = 0$.  Therefore, $\inner{v}{u} = \inner{z}{y}$, $\inner{v}{\sigma u} = \inner{z}{\sigma y}$, and $\inner{v}{\sigma^2 u} = \inner{z}{\sigma^2 y}$.  But since $y$ is of unit length, the same argument that led to~\eqref{eq:u-v} shows that the unique vector $z \in R_\sigma$ such that $\inner{z}{y} = -1$, $\inner{z}{\sigma y} = +1$, and $\inner{z}{\sigma^2 y} = 0$ is 
\begin{equation}
\label{eq:z}
z = \frac{2}{3} (\sigma y - y) \, . 
\end{equation}
Thus we have
\begin{equation}
\label{eq:perp}
u = y + u_\perp \quad \text{and} \quad v = z + v_\perp 
\end{equation}
where $y \in R_\sigma$ has unit length, $z \in R_\sigma$ is given by~\eqref{eq:z}, and $\inner{u_\perp}{v_\perp} = 0$.

This greatly reduces the dimensionality of our search, to choosing $y$ and finding $u_\perp$ and $v_\perp$.  We can reduce the search further by considering how $\ket{u}\bra{v}$ projects into the various trivial subspaces of $G$. As we did for $n=2$ above, it will be convenient to do this ``Fourier analytically,'' that is, by breaking $\rho \otimes \rho^*$ up into irreps under the $G^3$ conjugation action and using them to define a judicious choice of basis for $(\rho \otimes \rho^*)^{\otimes 3}$. While for larger groups this calculation becomes rather lengthy, note that because of our ansatz~\eqref{eq:family}, we only have $2n-1$ parameters to solve for. The conditions~\eqref{eq:uv-necessary}, resulting in the partial solution~\eqref{eq:perp}, reduce the number of parameters even further: there is a one-parameter family of unit vectors $y$ in $R_\sigma$, and $u_\perp$ and $v_\perp$ together have $(n-2)+(n-3)=2n-5$ free parameters.  As a result it suffices to use just a few of the projections onto irreps to reduce to a finite set of possible solutions.

\subsection{$n=3$: the standard representation of $S_4$}

For $n=3$, we let $\rho$ be the standard representation of $G=S_4$, which is of order $|S_4| = 24 = n^3-n$.  A convenient basis for this representation treats group elements as signed permutation matrices, whose determinants are the parities of the corresponding permutations.  For instance, 
\[
\rho((12)) = \begin{pmatrix}
0 & 1 & 0 \\
1 & 0 & 0 \\
0 & 0 & 1
\end{pmatrix} , \; 
\rho((23)) = \begin{pmatrix} 
1 & 0 & 0 \\
0 & 0 & 1 \\
0 & 1 & 0
\end{pmatrix} , \; 
\rho((34)) = \begin{pmatrix} 
0 & -1 & 0 \\
-1 & 0 & 0 \\
0 & 0 & 1 
\end{pmatrix} .
\]
For our order-3 element we take
\[
\sigma 
= \rho((123)) 
= \begin{pmatrix} 
0 & 0 & 1 \\
1 & 0 & 0 \\
0 & 1 & 0 
\end{pmatrix} . 
\]
These matrices permute the vertices of a tetrahedron,
\[
w_1 = (-1,+1,+1) \, , \quad 
w_2 = (+1,-1,+1) \, , \quad 
w_3 = (+1,+1,-1) \, , \quad 
w_4 = (-1,-1,-1) \, ,   
\]
according to the natural action of $S_4$ on a set of size 4.

As stated above, our ansatz~\eqref{eq:family} and the conditions~\eqref{eq:perp} implied by~\eqref{eq:uv-necessary} already reduce our search a great deal.  Specifically, we need to choose a unit vector $y$ in the plane $R_\sigma$ rotated by $\sigma$, giving us a free parameter $\theta$, as in Section~\ref{sec:strassen}.  This plane is perpendicular to $w_4$, so we have 
\begin{equation}
\label{eq:s4-perp}
u = y + aw_4
\quad \text{and} \quad
v = z + bw_4 \, .
\end{equation}
Moreover, $\inner{u_\perp}{v_\perp} = 0$, so either $a=0$ or $b=0$, giving two families with two free parameters.  Here we derive additional constraints using the Fourier technique we used for $n=2$.  These will suffice to reduce $u$ and $v$ to a finite number of possibilities, which indeed give valid matrix multiplication algorithms.

Let us denote the irreducible representations of $S_4$ by $\id$ (trivial), $\pi$ (sign), $\boxrep$ for the 2-dimensional irreducible representation, $\rho$ for the standard representation, and $\rho' = \rho \otimes \pi$ for the ``anti-standard'' representation. Using a little character theory, we find that $\rho \otimes \rho^* \cong \rho \otimes \rho \cong \id \oplus \boxrep \oplus \rho \oplus \rho'$.  Specifically, $\id, \boxrep, \rho$, and $\rho'$ consist respectively of the scalars, the diagonal matrices with trace zero, the symmetric matrices that are zero on the diagonal, and the antisymmetric matrices.  With some additional choices, this  decomposition into irreps yields the following orthogonal basis of $\rho \otimes \rho^*$: 
\begin{gather}
\id = \left\langle \id = \begin{pmatrix} 
1 & 0 & 0 \\
0 & 1 & 0 \\
0 & 0 & 1
\end{pmatrix}\right\rangle \; , \quad
\boxrep = \left\langle 
\boxrep_x = \begin{pmatrix}
1& 0 & 0 \\
0 & -1/2 & 0 \\
0 & 0 & -1/2
\end{pmatrix} \, , \;
\boxrep_y = \begin{pmatrix}
0 & 0 & 0 \\
0 & \sqrt{3}/2 & 0 \\
0 & 0 & -\sqrt{3}/2
\end{pmatrix}
\right\rangle \nonumber \\
\rho = \left\langle
s_1 = \begin{pmatrix}
0 & 0 & 0 \\
0 & 0 & 1 \\
0 & 1 & 0 
\end{pmatrix} \, , \; 
s_2 = \begin{pmatrix}
0 & 0 & 1 \\
0 & 0 & 0 \\
1 & 0 & 0
\end{pmatrix} \, , \;
s_3 = \begin{pmatrix}
0 & 1 & 0 \\
1 & 0 & 0 \\
0 & 0 & 0 
\end{pmatrix}
\right\rangle \nonumber \\
\rho' = \left\langle
a_1 = \begin{pmatrix}
0 & 0 & 0 \\
0 & 0 & 1 \\
0 & -1 & 0
\end{pmatrix} \, , \;
a_2 = \begin{pmatrix}
0 & 0 & -1 \\
0 & 0 & 0 \\
1 & 0 & 0
\end{pmatrix} \, , \;
a_3 = \begin{pmatrix}
0 & 1 & 0 \\
-1 & 0 & 0 \\
0 & 0 & 0
\end{pmatrix}
\right\rangle
\label{eq:s4-basis}
\end{gather}
We chose $s_1, s_2, s_3$ so that they correspond to the original basis $e_1 = (1,0,0), e_2 = (0,1,0), e_3 = (0,0,1)$ of $\rho$. Similarly, since $\rho' \cong \rho \otimes \pi$, we chose $a_1, a_2, a_3$ to be compatible with this mapping in the basis of $\rho$ that we started with.  

The Klein four-group $K_4 \cong \Z_2 \times \Z_2$, which appears in the derived series of $S_4$ and is generated by $(12)(34)$ and $(13)(24)$, consists of the diagonal matrices with an even number of $-1$s.  Since these commute with $\boxrep_x$ and $\boxrep_y$ and we are applying the conjugation action to $\rho \otimes \rho^*$, we recover the fact that $K_4$ is the kernel of $\boxrep$, and that $\boxrep$ is the standard irrep of the quotient $S_4/K_4 \cong S_3$.  We chose $\boxrep_x$ and $\boxrep_y$ so that they correspond with same basis vectors $(1,0), (0,1)$ of this irrep as in Section~\ref{sec:strassen}, and in particular of the copy of $S_3 \subset S_4$ that fixes $w_4$ and rotates and reflects $R_\sigma$.  

To derive our additional equations, we will use~\eqref{eq:family} to compute $\inner{\T}{A \otimes B \otimes C} = \tr ABC$ for certain triples $A, B, C$ of basis elements from~\eqref{eq:s4-basis}.  Recall that since $\T$ is fixed under the diagonal conjugation action of $S_4$, it can only have a nonzero projection into triples $\mu \otimes \nu \otimes \lambda$ that contain a copy of the trivial irrep of the diagonal action; thus only basis elements drawn from such triples are informative.

One such triple is $\id \otimes \rho \otimes \rho$, which contains a copy of the trivial irrep spanned by 
\[
\id \otimes (s_1 \otimes s_1 + s_2 \otimes s_2 + s_3 \otimes s_3) \, . 
\]
Computing $\inner{\T}{\id \otimes s_1 \otimes s_1} = \tr s_1^2$ using~\eqref{eq:family} and recalling $\inner{v}{u} = -1$ from~\eqref{eq:uv-necessary} yields
\[
2 = 0 + \inner{v}{u} \sum_{g \in S_4} \bra{v} s_1^{g} \ket{u} \bra{v} s_1^{g \sigma} \ket{u} 
= - \sum_{g \in S_4} \bra{v} s_1^{g} \ket{u} \bra{v} s_1^{g \sigma} \ket{u} \, ,
\]
where we adopt the shorthand $s^g = \rho(g)^\dagger s \rho(g)$ and (abusing notation slightly) $s^{g \sigma} = \sigma^\dagger \rho(g)^\dagger s \rho(g) \sigma$.  Note that the $0$ comes from $\inner{\id^{\otimes 3}}{\id \otimes s_1 \otimes s_1} = (\tr \id)(\tr s_1)^2$.

Since we chose $s_1, s_2, s_3$ so that $s_i$ corresponds with $e_i$ in our original basis for $\rho$, we have $s_1^g 
= (-1)^{\eps_i} s_i$ where the first row of $\rho(g)$ is $(-1)^{\eps_i} e_i$. Similarly, $s_1^{g \sigma} 
= (-1)^{\eps_j} s_j$ where the first row of $\rho(g) \sigma$ is $(-1)^{\eps_j} e_j$.  However, since $\sigma$ is a permutation matrix with no negative entries, $\eps_j = \eps_i$ and the signs cancel out.  Moreover, multiplying $\rho(g)$ by $\sigma$ on the right permutes the $e_j$: that is, $j=(1,2,3)^{-1} i \ne i$.  As a result, each distinct pair $i, j$ of indices appears in $8$ of the $24$ terms in the sum, and we are left with 
\begin{equation}
\bra{v} s_1 \ket{u} \bra{v} s_2 \ket{u} 
+ \bra{v} s_2 \ket{u} \bra{v} s_3 \ket{u} 
+ \bra{v} s_3 \ket{u} \bra{v} s_1 \ket{u} 
= -\frac{1}{4} \, . \label{eq:4std}
\end{equation}

We reason similarly in $\id \otimes \rho' \otimes \rho'$, where the trivial irrep is spanned by 
\[
\id \otimes (a_1 \otimes a_1 + a_2 \otimes a_2 + a_3 \otimes a_3) \, . 
\]
Computing $\inner{\T}{\id \otimes a_1 \otimes a_1} = \tr a_1^2$ using~\eqref{eq:family} yields
\[
-2 = 0 + \inner{v}{u} \sum_{g \in S_4} \bra{v} a_1^{g} \ket{u} \bra{v} a_1^{g \sigma} \ket{u} 
= -\sum_{g \in S_4} \bra{v} a_1^{g} \ket{u} \bra{v} a_1^{g \sigma} \ket{u} \, .
\]
Recall that $\rho'(g) = \pi(g) \rho(g)$ where $\pi(g) = \pm 1$ is the parity.  Since $(123)$ is even, $g$ and $g(123)$ have the same parity, so the parity disappears from each term in the sum.  Following the same reasoning then gives
\begin{equation}
\bra{v} a_1 \ket{u} \bra{v} a_{2} \ket{u} 
+ \bra{v} a_2 \ket{u} \bra{v} a_3 \ket{u} 
+ \bra{v} a_3 \ket{u} \bra{v} a_1 \ket{u} 
= \frac{1}{4} \, . \label{eq:4anti-std}
\end{equation}

We will derive one more constraint along these lines.  Analogous to~\eqref{eq:rhorhorho}, $\boxrep \otimes \boxrep \otimes \boxrep$ contains a copy of the trivial irrep spanned by
\[
\boxrep_x \otimes \boxrep_x \otimes \boxrep_x 
- \boxrep_x \otimes \boxrep_y \otimes \boxrep_y 
- \boxrep_y \otimes \boxrep_x \otimes \boxrep_y 
- \boxrep_y \otimes \boxrep_y \otimes \boxrep_x \, . 
\]
We compute $\inner{\T}{\boxrep_x \otimes \boxrep_x \otimes \boxrep_x} = \tr \boxrep_x^3$, giving
\[
\frac{3}{4} = 0 + \sum_{g \in S_4} 
\bra{v} \boxrep_x^{g} \ket{u} 
\bra{v} \boxrep_x^{g \sigma} \ket{u} 
\bra{v} \boxrep_x^{g \sigma^2} \ket{u} \, . 
\]
Recall that $\boxrep$ is the standard irrep of $S_4/K_4 \cong S_3$.  Moreover, $\beta_x$ is stabilized by the transposition $\rho((23))$, so it has an orbit of size $3$.  Specifically, all its conjugates are either $\boxrep_x$, $\boxrep_x^\sigma$, or $\boxrep_x^{\sigma^2}$, and these differ only in which diagonal entry is $1$.  Each of the $24$ terms in the sum contains all three of these, so we have
\begin{equation}
\label{eq:4box}
\bra{v} \boxrep_x \ket{u} 
\bra{v} \boxrep_x^{\sigma} \ket{u} 
\bra{v} \boxrep_x^{\sigma^2} \ket{u} 
= \frac{1}{32} \, . 
\end{equation}

With~\eqref{eq:4std}, \eqref{eq:4anti-std}, and~\eqref{eq:4box} in hand, we can further restrict the two-parameter families described by~\eqref{eq:s4-perp} until only valid decompositions of the matrix multiplication tensor remain.  First we write a one-parameter family of unit vectors $y \in R_\sigma$, 
\[
y = y(\theta) = -\sqrt{ \frac{2}{3} } \big( \! \cos \theta, \cos (\theta-2\pi/3), \cos (\theta-4\pi/3) \big) \, . 
\]
in which case
\[
z = \frac{2}{3} (\sigma y - y) 
= -\frac{2}{\sqrt{3}} \,y(\theta+5 \pi/6) \, . 
\]
Since either $a=0$ or $b=0$ is zero in~\eqref{eq:s4-perp}, we start with 
\begin{equation}
\label{eq:family1}
u = y + a w_4 
\quad \text{and} \quad 
v = z \, . 
\end{equation}
Plugging this into~\eqref{eq:4anti-std} yields a constraint on $a$ which is independent of $\theta$,
\[
\frac{1}{3} - 2a^2 = \frac{1}{4}
\quad \text{so} \quad
a = \pm \frac{1}{2\sqrt{6}} \, . 
\]
If we set $a = -1/(2\sqrt{6})$, then~\eqref{eq:4std} becomes
\[
1 - \cos 3\theta - \sqrt{3} \sin 3\theta = 0 \, . 
\]
This implies that $\theta = (2\pi/3)k$ or $(2\pi/3)(k+1/3)$ for integer $k$.  The second of these is eliminated by~\eqref{eq:4box}, which becomes
\[
18 \cos 3\theta + 7 \sqrt{3} \sin 3\theta - 8 \sqrt{3} \sin 6\theta = 18 \, . 
\]
Thus $\theta$ is an integer multiple of $2\pi/3$.  In particular, with $\theta=0$ we have 
\[
y = \frac{1}{2\sqrt{6}} (2 w_1 - w_2 - w_3) \, ,
\]
and setting $a=-1/(2 \sqrt{6})$ and using $w_4 = -(w_1+w_2+w_3)$ gives
\[
u = y - \frac{1}{2\sqrt{6}} w_4 
= \frac{1}{2} \sqrt{\frac{3}{2}} \,w_1 
= \frac{1}{2} \sqrt{\frac{3}{2}} (-1,+1,+1) 
\quad \text{and} \quad 
v = z = \frac{1}{\sqrt{6}} (w_2-w_1) 
= \sqrt{\frac{2}{3}} (1,-1,0) \, . 
\]
This gives an algorithm for $n=3$ matrix multiplication formed from the rank-1 matrix
\begin{equation}
\label{eq:first-family}
m 
= \ket{u} \bra{v} 
= \frac{1}{2} 
\begin{pmatrix}
-1 & +1 & 0 \\
+1 & -1 & 0 \\
+1 & -1 & 0 
\end{pmatrix} \, . 
\end{equation}
Taking~\eqref{eq:family1} with $a=1/(2 \sqrt{6})$ gives solutions whenever $\theta = (2\pi/3)(k+1/2)$, leading to alternate rank-1 matrices which are conjugate to this one.

We also explored the family where $u=y$ and $v=z+b w_4$.  In this case, \eqref{eq:4anti-std} implies that $b = \pm 1/(3 \sqrt{2})$.  Setting $b=-1/(3\sqrt{2})$ yields solutions at $\theta = (2\pi/3)(k+3/4)$, and setting $b=1/(3 \sqrt{2})$ yields solutions at $\theta = (2\pi/3)(k+1/4)$.  The resulting rank-1 matrices are conjugates of 
\begin{equation}
\label{eq:second-family}
m' 
= \frac{1}{2} 
\begin{pmatrix}
-1 & +1 & +1 \\
0 & 0 & 0 \\
+1 & -1 & -1 
\end{pmatrix} \, . 
\end{equation}
These are conjugates of the transpose of the first family of solutions~\eqref{eq:first-family}.  Specifically,
\[
m' = \tau m^\dagger \tau 
\]
where $\tau$ is a transposition; in this case, $\tau = \rho((23))$.  Indeed, an easy exercise shows that for any $m$ which gives a valid solution to~\eqref{eq:family}, $\tau^{-1} m^\dagger \tau$ does as well, as long as $\tau^{-1} \sigma \tau = \sigma^{-1}$.  We leave this to the reader.

\subsection{$n=4$: the standard representation of $A_5$}

For larger $n$, the pattern continues. Rather than writing out the decomposition into irreps explicitly, we simply summarize the results of our calculations.

For $n=4$, we let $\rho$ be the standard representation of $A_5$, which is of order $|A_5| = 60 = n^3-n$.  We construct unit vectors $w_1, \ldots, w_5$ at the corners of a simplex which are permuted by $A_5$.  The two-dimensional subspace $R_\sigma$ rotated by $\sigma = \rho((123))$ is perpendicular to $w_4$ and $w_5$, and in particular to $w_1+w_2+w_3 = -(w_4+w_5)$.  In this plane we define
\[
y = \sqrt{ \frac{2}{15} } \,(2 w_1 - w_2 - w_3) 
\quad \text{and} \quad
z = \frac{2}{3} (\sigma y - y) 
= 2 \sqrt{ \frac{2}{15} } \,(w_2 - w_1) \, .
\]
Using a mixture of numerical search and computational algebra, we found a solution of the form~\eqref{eq:perp}, which we do not claim is unique:
\[
u = y + \sqrt{ \frac{2}{15} } \,(w_1+w_2+w_3) = \sqrt{ \frac{6}{5} } \,w_1
\quad \text{and} \quad
v = z 
= 2 \sqrt{ \frac{2}{15} } \,(w_2 - w_1) \, .
\]

Unlike the previous case, this representation is not monomial, so it's difficult to find a basis which is both elegant and explicit.  One basis, obtained by projecting the $5$ unit vectors in $\R^5$ along Fourier basis vectors with nonzero frequencies, places the corners of the simplex at
\begin{gather*}
w_1 = \frac{1}{\sqrt{2}} \left( 1, 0, 1, 0 \right) \\
w_2 = \frac{1}{4} \left( \frac{\sqrt{5}-1}{\sqrt{2}}, \sqrt{5+\sqrt{5}}, -\frac{\sqrt{5}+1}{\sqrt{2}}, \sqrt{5-\sqrt{5}} \right) \\
w_3 = \frac{1}{4} \left( -\frac{\sqrt{5}+1}{\sqrt{2}}, \sqrt{5-\sqrt{5}}, \frac{\sqrt{5}-1}{\sqrt{2}}, -\sqrt{5+\sqrt{5}} \right) \\
w_4 = \frac{1}{4} \left( -\frac{\sqrt{5}+1}{\sqrt{2}}, -\sqrt{5-\sqrt{5}}, \frac{\sqrt{5}-1}{\sqrt{2}}, \sqrt{5+\sqrt{5}} \right) \\
w_5 = \frac{1}{4} \left( \frac{\sqrt{5}-1}{\sqrt{2}}, -\sqrt{5+\sqrt{5}}, -\frac{\sqrt{5}+1}{\sqrt{2}}, -\sqrt{5-\sqrt{5}} \right) \, . 
\end{gather*}
In that case we have 
\[
u = \sqrt{ \frac{3}{5} } \left( 1, 0, 1, 0 \right) 
\quad \text{and} \quad
v = \frac{1}{\sqrt{3}} 
\left( 
-\frac{\sqrt{5} - 1}{2},
\frac{1}{2} \sqrt{\frac{\left( 3+\sqrt{5} \right)\left( 5-\sqrt{5} \right)}{5}},
-\frac{\sqrt{5}+1}{2},
\sqrt{\frac{5-\sqrt{5}}{10}} 
\right) \, , 
\]
and 
\[
m = \ket{u} \bra{v}
= \frac{1}{\sqrt{5}}
\begin{pmatrix}
-\frac{\sqrt{5} - 1}{2} &
\frac{1}{2} \sqrt{\frac{\left( 3+\sqrt{5} \right)\left( 5-\sqrt{5} \right)}{5}} &
-\frac{\sqrt{5}+1}{2} &
\sqrt{\frac{5-\sqrt{5}}{10}} \\
0 & 0 & 0 & 0 \\
-\frac{\sqrt{5} - 1}{2} &
\frac{1}{2} \sqrt{\frac{\left( 3+\sqrt{5} \right)\left( 5-\sqrt{5} \right)}{5}} &
-\frac{\sqrt{5}+1}{2} &
\sqrt{\frac{5-\sqrt{5}}{10}} \\
0 & 0 & 0 & 0 
\end{pmatrix} \, .
\]

\subsection{$n=5$: an irrep of $S_5$}

For $n=5$, we again look for a group of size $n^3-n = 120$.  A natural choice is $S_5$, which has two five-dimensional irreps which differ by the parity.  We focus on $\rho = {\scriptsize \Yvcentermath1 \yng(3,2)}$, which is one of the irreducible components of the natural permutation action on subsets $U \subset \{1,2,3,4,5\}=[5]$ with $|U|=2$.  It is spanned by the set of functions $f\colon \binom{[5]}{2} \to \R$ such that, for all $i \in \{1,2,3,4,5\}$, $\sum_{U \ni i} f(U) = 0$.

If we express $\rho$ using the Young orthogonal basis~\cite{young-orthogonal}, with basis vectors corresponding to the Young tableaux 

\[
\young(123,45) \; , \quad
\young(124,35) \; , \quad
\young(125,34) \; , \quad
\young(134,25) \; , \quad
\young(145,23) \; , 
\]
\smallskip 

\noindent
then we have the order-3 element
\[
\sigma = \rho((123))
= \begin{pmatrix} 
 1 & 0 & 0 & 0 & 0 \\
 0 & -\frac{1}{2} & 0 & \frac{\sqrt{3}}{2} & 0 \\
 0 & 0 & -\frac{1}{2} & 0 & \frac{\sqrt{3}}{2} \\
 0 & -\frac{\sqrt{3}}{2} & 0 & -\frac{1}{2} & 0 \\
 0 & 0 & -\frac{\sqrt{3}}{2} & 0 & -\frac{1}{2} 
 \end{pmatrix} \, . 
\]
Note that unlike our previous constructions, $\sigma$ rotates \emph{two} distinct two-dimensional subspaces rather than one.  This gives $u$ and $v$ additional degrees of freedom, meaning that~\eqref{eq:perp} might not apply.  However, we can leap directly to a solution inspired by an apparent pattern in the examples above.

As the reader may have noticed, in each case above, we had a set $W$ of vectors $w_1, \ldots, w_k$ which form an orbit under the group, and found a decomposition based on the rank-1 matrix $m=\ket{u} \bra{v}$ where $u$ and $v$ are proportional to $w_1$ and $w_2-w_1$ respectively.  In $n=2, 3, 4$ this orbit had size $|W|=n+1$, with the $w_i$ pointing to the corners of a simplex which the group permutes.  With $G=S_3$, $S_4$, and $A_5$, each $w_i$ had a stabilizer subgroup $H \cong \Z_2$, $S_3$, and $A_4$ respectively.  

The group $S_5$ does not have any orbits of size $6$ in $\rho$.  However, its subgroup $A_5$ has an orbit $\{w_1,\ldots,w_6\}$ which is sent to $\{-w_1,\ldots,-w_6\}$ by the odd permutations, so $S_5$ has an orbit of size $12$ of the form $\{ \pm w_1, \ldots, \pm w_6 \}$. This orbit has a stabilizer of size 20 which is determined up to conjugacy in $S_5$, and there is a unique conjugacy class of subgroups of $S_5$ of order 20. Namely, identify $\{1,2,3,4,5\}$ with $\F_5$ arbitrarily, and let $H$ be the subgroup consisting of $\F_5$-affine functions $x \mapsto ax+b$ where $a \in \F_5^*$ and $b \in \F_5$; note that $|H|=20$.  Then restricting $S_5$ to $H$ breaks $\rho$ into a 4-dimensional irrep of $H$ and a copy of the sign representation, and each one-dimensional subspace spanned by $\pm w_i$ is fixed up to a sign by $H$ or one of its conjugates. (In particular, each $w_i$ is defined by its stabilizer, in the same way that $\T$ itself is.)

Then if we take
\[
w_1 = \frac{1}{\sqrt{5}} \left( 1, \sqrt{2}, 0, 0, \sqrt{2} \right) 
\quad \text{and} \quad
w_2 = \sigma w_1 
= \frac{1}{\sqrt{10}} \left( \sqrt{2}, -1, \sqrt{3}, -\sqrt{3}, -1 \right) \, , 
\]
we find using computational algebra that 
\[
u = w_1 
\quad \text{and} \quad
v = \frac{5}{6} (w_2-w_1)
= \sqrt{ \frac{5}{6} } \left( 0, -\frac{\sqrt{3}}{2}, \frac{1}{2}, -\frac{1}{2}, -\frac{\sqrt{3}}{2} \right) 
\]
indeed gives a valid decomposition of the matrix multiplication tensor.

\section{Lattices and a construction for general $n$}

In all the constructions we have found, the inner products of the $w_i \in W$ are rational: specifically, $\inner{w_i}{w_j} = -1/n$ if we normalize the $w_i$ to have unit length.  It follows that $W$ generates a lattice, and we can take $u$ and $v$ to be elements of this lattice up to an overall scale factor.  This strongly suggests to us that good constructions for larger $n$ might involve lattices whose symmetry group contains $G$.  

In fact, after rescaling $u$ and $v$, all our constructions can be written as follows, where $w_1,\ldots,w_{n+1}$ form a frame of unit vectors that point to the corners of a simplex:
\begin{equation}
\label{eq:lattice}
\T_n = \id^{\otimes 3} 
+ \left( \frac{n}{n+1} \right)^3 \sum_{\mbox{$i,j,k$ distinct}} 
\ket{w_i}\bra{w_j-w_i} \otimes
\ket{w_j}\bra{w_k-w_j} \otimes
\ket{w_k}\bra{w_i-w_k} \, . 
\end{equation}
This sum over all $(n+1)n(n-1)=n^3-n$ distinct ordered triples occurs whenever the group $G$ acts 3-transitively on the $w_i$.  This is true for $S_3$ in $n=2$, $S_4$ in $n=3$, $A_5$ in $n=4$, and (projectively) $S_5$ in $n=6$.  (Indeed, the 3-transitive permutation groups have been completely classified, see, e.\,g., \cite{}.) Note that this last action relies on the unusual embedding of $S_5$ in $S_6$, which we can obtain by taking a standard embedding (say, the copy $S_5 \subset S_6$ that fixes the 6th element) and applying the outer automorphism of $S_6$.

There is a simple proof that~\eqref{eq:lattice} gives a valid matrix multiplication algorithm for any $n$.  First note that we can sum over all $(n+1)^3$ triples $i,j,k$, since if any of these are equal the summand is zero.  We also have 
\[
\sum_i \ket{w_i} = 0 
\quad \text{and} \quad
\frac{n}{n+1} \sum_i \ket{w_i} \bra{w_i} = \id \, .
\]
Then 
\begin{align}
&\left( \frac{n}{n+1} \right)^3 \sum_{i,j,k}
\ket{w_i}\bra{w_j-w_i} \otimes
\ket{w_j}\bra{w_k-w_j} \otimes
\ket{w_k}\bra{w_i-w_k} \nonumber \\
&= \left( \frac{n}{n+1} \right)^3 \left[ \sum_{i,j,k}
\ket{w_i}\bra{w_j} \otimes
\ket{w_j}\bra{w_k} \otimes
\ket{w_k}\bra{w_i} \right. \nonumber \\
&\quad - \sum_{i,j,k} \Big[ 
\ket{w_i}\bra{w_i} \otimes
\ket{w_j}\bra{w_k} \otimes
\ket{w_k}\bra{w_i} 
+ \ket{w_i}\bra{w_j} \otimes
\ket{w_j}\bra{w_j} \otimes
\ket{w_k}\bra{w_i} 
+ \ket{w_i}\bra{w_j} \otimes
\ket{w_j}\bra{w_k} \otimes
\ket{w_k}\bra{w_k} \Big] \label{eq:line3} \\
&\quad + \sum_{i,j,k} \Big[ 
\ket{w_i}\bra{w_j} \otimes
\ket{w_j}\bra{w_j} \otimes
\ket{w_k}\bra{w_k} 
+ \ket{w_i}\bra{w_i} \otimes
\ket{w_j}\bra{w_k} \otimes
\ket{w_k}\bra{w_k} 
+ \ket{w_i}\bra{w_i} \otimes
\ket{w_j}\bra{w_j} \otimes
\ket{w_k}\bra{w_i} \Big] \label{eq:line4} \\
&\quad - \left. \sum_{i,j,k}
\ket{w_i}\bra{w_i} \otimes
\ket{w_j}\bra{w_j} \otimes
\ket{w_k}\bra{w_k} \right] \nonumber \\
&= \left( \frac{n}{n+1} \right)^3 \sum_{i,j,k}
\ket{w_i}\bra{w_j} \otimes
\ket{w_j}\bra{w_k} \otimes
\ket{w_k}\bra{w_i} 
- \left( \frac{n}{n+1} \sum_i \ket{w_i} \bra{w_i} \right)^{\otimes 3} \label{eq:frameworks} \\
&= \T_n - \id^{\otimes 3} \, . \nonumber 
\end{align}
The mixed terms in lines~\eqref{eq:line3} and~\eqref{eq:line4} disappear because each product has an index that appears only once, and summing over that index gives zero.  To see that the first term in~\eqref{eq:frameworks} is $\T_n$, first note that since $W$ is an orbit of an irreducible group action, such as the standard irrep of $G=S_{n+1}$, this term has the $G^3$ symmetries of matrix multiplication discussed in Section~\ref{sec:strassen}.  This shows that it is a constant times $\T_n$, and we can check that this constant is $1$ by taking the inner product:
\begin{gather*}
\left\langle \T_n \left| \left( \frac{n}{n+1} \right)^3 \sum_{i,j,k}
\ket{w_i}\bra{w_j} \otimes
\ket{w_j}\bra{w_k} \otimes
\ket{w_k}\bra{w_i} \right. \right\rangle 
= \left( \frac{n}{n+1} \right)^3 \sum_{i,j,k} \tr \big( \ket{w_i} \inner{w_j}{w_j} \inner{w_k}{w_k} \bra{w_i} \big) \\
= n^3 = \inner{\T_n}{\T_n} \, . 
\end{gather*}
This gives a very compact and transparent proof that Strassen's algorithm is a valid decomposition of $\T_2$, and shows that for all $n$ there is a decomposition of $\T_n$ of tensor rank $n^3 - n + 1$.

\section*{Acknowledgments} Parts of this project were inspired in 2015 by discussions with Jonah Blasiak, Thomas Church, Henry Cohn, and Chris Umans, via a collaboration funded by the AIM SQuaRE program, with an additional visit hosted by the Santa Fe Institute.  J.A.G. was funded by an Omidyar Fellowship from the Santa Fe Institute during this work, and C.M. was funded partly by the John Templeton Foundation.  C.M. also thanks \'Ecole Normale Sup\'erieure for providing a visiting position during which some of this work was carried out.

\bibliographystyle{alphaurl}
\bibliography{strassen-fourier}

\end{document}